\input stromlo

\title Dynamics of Triaxial Stellar Systems

\shorttitle Triaxial Stellar Systems

\author David Merritt

\shortauthor D Merritt

\affil Rutgers University, New Brunswick, NJ

\abstract 
Recent work on the dynamics of triaxial stellar systems is reviewed.
The motion of boxlike orbits in realistic triaxial potentials is 
generically stochastic.
The degree to which the stochasticity manifests itself in the 
dynamics depends on the chaotic mixing timescale, which is 
a small multiple of the crossing time in triaxial models with steep 
cusps or massive central singularities.
Low-luminosity ellipticals, which have the steepest cusps and the
shortest dynamical times, are less likely than bright ellipticals 
to have strongly triaxial shapes.
The observational evidence for triaxiality is reviewed; 
departures from axisymmetry in early-type galaxies are often
found to be associated with evidence of recent interactions or with the
presence of a bar.

\section Introduction

Triaxiality began as a plausibility argument (Binney 1978).
Since 1975, elliptical galaxies had been known to be rotating 
too slowly for their shapes to be due to centrifugal
flattening.
Hence there was no compelling reason for them to be 
oblate, and triaxiality seemed a natural alternative.
The discovery that orbits in triaxial potentials were
often regular, i.e. non-chaotic, coupled with the 
seeming ease with which self-consistent triaxial models 
could be constructed on the computer (Schwarzschild 1979, 1982) 
lent further support to the hypothesis.
The geometrically simpler alternative -- that elliptical galaxies are 
axisymmetric, and that their slow rotation is due to the 
cancellation of angular momentum by stars orbiting in 
opposite directions about the symmetry axis -- seemed contrived.
Furthermore a growing body of observational evidence 
suggested that many early-type galaxies were not axisymmetric.

The case for triaxiality is perhaps less compelling now than it
was fifteen years ago.
Box orbits, the ``backbone'' of triaxial galaxies, require for 
their existence a constant-density core where the motion is 
nearly harmonic.
We now know that nature never provides elliptical galaxies with 
such cores; indeed, massive central singularities may be the norm
(Kormendy \& Richstsone 1995).
It is also clear that nature is capable of making stellar 
systems that are both pressure-supported and close to axisymmetric 
(Rubin et al. 1992; Merrifield \& Kuijken 1994).
The link between velocity anisotropy and triaxiality is 
thus weakened.
Finally, while some early-type galaxies are definitely not 
axisymmetric, many of these galaxies may be barred S0's
or systems that are not fully relaxed -- quite 
different from the original conceptual model of stationary, 
nested ellipsoids.

In triaxial models that resemble real elliptical galaxies, most of the 
boxlike orbits are stochastic, respecting only the energy 
integral.
The potential of stochastic orbits to induce evolution of 
the global shapes of elliptical galaxies has long been recognized 
(Norman et al. 1985; Gerhard \& Binney 1985), but recent thinking about this 
problem has been sharpened by recasting it in terms of ``chaotic mixing,'' 
the mechanism by which an ensemble of points in stochastic phase 
space relaxes to a steady state.
In a galaxy where the chaotic mixing time is short compared to
a Hubble time, nature does not have the freedom to assign 
arbitrary densities to different parts of stochastic phase space,
any more than it can place all of the gas molecules in one corner of a room.
The resulting loss of freedom makes it more difficult to 
arrange stars into self-consistent triaxial equilibria.

\section Cores and Cusps

High resolution observations of elliptical galaxies by a number
of groups have consistently shown that nuclear density profiles
continue to rise into the smallest observable radius.
The non-existence of constant-density cores might 
have been recognized even before the era of HST.
For instance, in M87, a prototypical ``core'' galaxy, the 
surface brighntness measurements of Young et al. (1978) imply
a deprojected luminosity density that 
rises as a power-law inside of $\sim 10''$, the nominal core 
radius (e.g. Richstone \& Tremaine 1985, Fig. 1).
Galaxies like M87 appear to have cores because of an optical 
illusion associated with projection onto the plane of the sky.
A density profile that varies as $r^{-\gamma}$ at small radii 
generates a power-law cusp in projection only if $\gamma>1$.
When $\gamma=1$, the surface brightness exhibits a 
curving, logarithmically-divergent central profile 
(e.g. Dehnen 1993, Fig. 1), and for $\gamma<1$ the central 
surface brightness is finite.
The observed brightness profile of a galaxy like M87, which has 
$\gamma\approx 0.8$, differs only subtly from that of a galaxy with an 
isothermal core.

\figureps[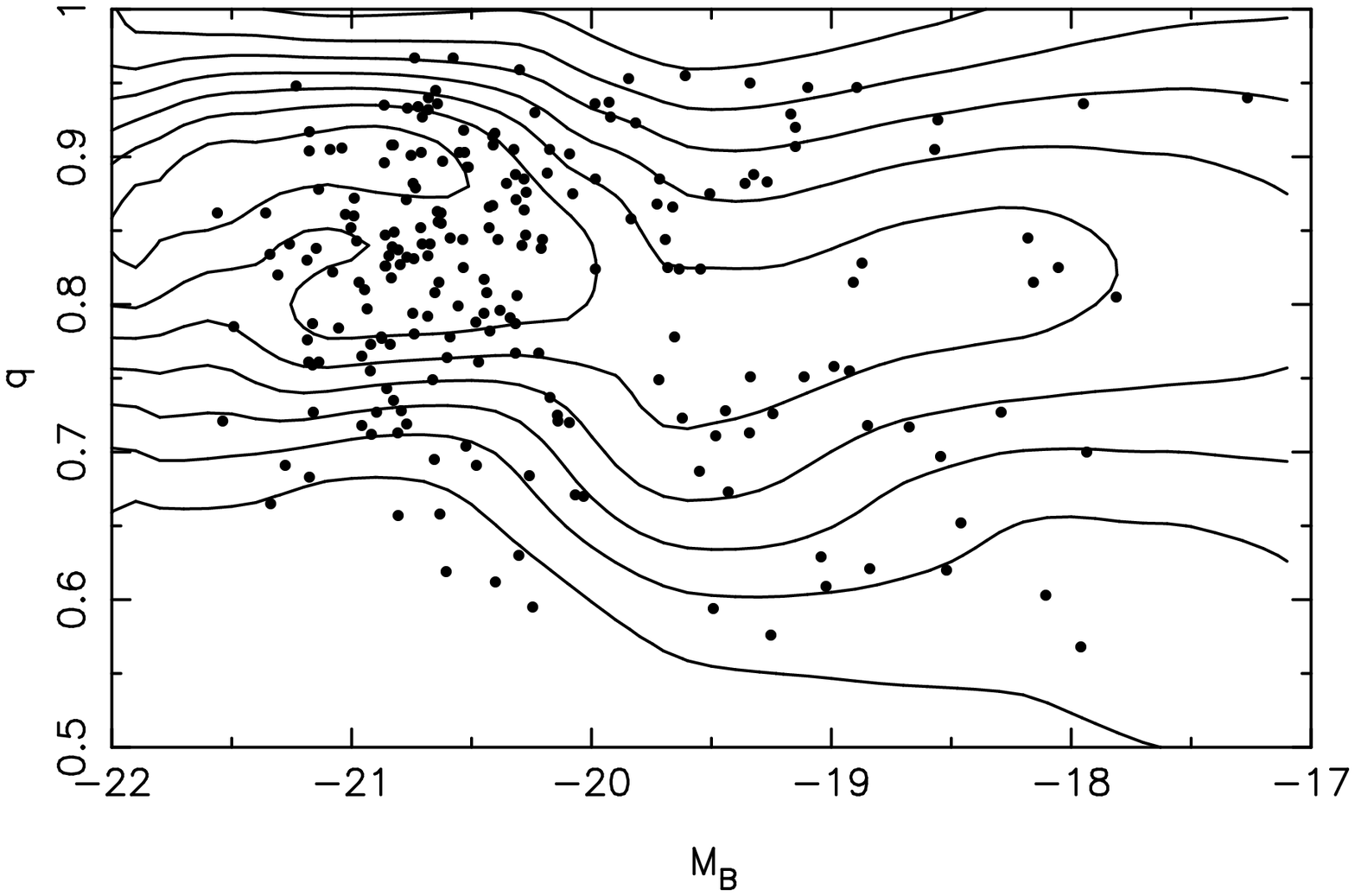,.6\hsize] 1. Hubble-type distribution of
elliptical galaxies  as a function of intrinsic luminosity $M_B$
(Tremblay \& Merritt 1996); $q$ is the apparent short-to-long 
axis ratio.
Contours show the frequency function of Hubble types $N(q)$, normalized 
to unit area at each $M_B$.
There is a striking change in the shape distribution at $M_B\approx 
-20$.

Thus even if galaxies were distributed uniformly over 
$\gamma$, their surface brightness profiles would appear to fall
naturally into one of two categories: the ``cores'' 
($\gamma\leq 1$) and the ``power-laws'' ($\gamma>1$).
Just such a characterization was adopted based on the first surface 
brightness measurements from HST (Ferrarese et al. 1994; Lauer et al. 1995).
The suggestion of Merritt \& Fridman (1995) that {\it all} elliptical 
galaxy nuclei might have power-law profiles in the {\it space} density -- 
differing only in the small-radius slope $\gamma$ -- was beautifully 
confirmed by Gebhardt et al. (1996), who used a nonparametric algorithm to 
deproject the HST data from a large sample of ellipticals.
Unfortunately, as often happens, the nomenclature has remained 
frozen and one still hears talk of the ``two types'' of surface 
brightness profile (e.g. Lauer, this volume).
Here we adopt a more neutral terminology: galaxies with 
deprojected profiles having $\gamma\leq 1$ are ``weak-cusp'' 
galaxies, while values of $\gamma$ greater than $1$ define a
``strong cusp.''

Even this dichotomy might seem artificial, since it is only
through a trick of projection that the value $\gamma=1$ appears
to be special. 
However there is a more fundamental reason for making the division at 
$\gamma=1$.
A central density that increases more rapidly than $r^{-1}$ 
implies a divergent central force.
For instance, a spherical galaxy with Dehnen's (1993) density 
law
$$
\rho(r) = {(3-\gamma)M\over 4\pi a^3}\left({r\over a}\right)^{-\gamma}
\left(1+{r\over a}\right)^{-(4-\gamma)}
$$
has a gravitational force
$$
-{\partial\Phi\over\partial r} = -{GM\over a^2}\left({r\over 
a}\right)^{1-\gamma}\left(1+{r\over a}\right)^{\gamma-3}.
$$
We might expect the character of the ``centrophilic'' orbits, like 
the boxes, to change radically as $\gamma$ is increased past 
$1$.
This expectation turns out to be correct, as discussed below.
Furthermore there is some indication that the shapes of 
elliptical galaxies with weak cusps are systematically different 
than those with strong cusps.
Gebhardt et al. (1996) find that $\gamma\geqsim 1$ for faint 
elliptical galaxies, $M_V\geqsim -20$, while $\gamma\leqsim 1$ 
for brighter ellipticals.
As Figure 1 shows, roughly the same absolute magnitude also neatly 
divides elliptical galaxies into two groups with very different 
distributions of apparent shapes.
This  difference may be due in part to the different behavior of 
boxlike orbits in galaxies with weak and strong cusps.

\section Regular Orbits

The only global integral of the motion in a generic triaxial potential is 
the energy $E=v^2/2 + \Phi$.
However we expect extra integrals to exist in the vicinity of 
stable, periodic orbits.
Periodic orbits fill phase space densely, but many of them 
are unstable, and furthermore the volume of regular phase space 
associated with a periodic orbit declines rapidly with 
the order of the resonance.
The most important perodic orbits are therefore those associated with 
the lowest-order stable resonances.

Planar, 1:1 closed orbits -- circular orbits in the axisymmetric 
geometry -- exist in a wide variety of triaxial potentials; they 
disappear only near the centers of triaxial potentials with 
cores (Merritt \& de Zeeuw 1982) where the motion is nearly 
harmonic and the orbital frequencies incommensurate.
The 1:1 orbits circling the long and short axes of a triaxial 
model are generally stable (Heiligman \& Schwarzschild 1979), 
and perturbations of these orbits produce the long- and short-axis 
tube families (Kuzmin 1973; de Zeeuw 1985).
The fact that tube orbits are present outside the core in almost all
triaxial potentials justifies the use of highly simplified models,
like the Perfect Ellipsoid, to model the rotational velocity fields far 
from the centers of real elliptical galaxies (Statler 1991).

The axial orbits constitute a second major class of periodic 
orbit.
The long-axis orbit, when stable, generates box orbits, which 
are uniquely associated with the triaxial geometry.
In models with constant-density cores, the long-axis orbit remains 
stable from the center out to large radii; it first becomes 
unstable when the frequency of oscillation along the long axis falls 
to $1/2$ the average oscillation frequency in the direction of 
the short or intermediate axis.
A bifurcation then occurs, with the 2:1 ``x-z banana'' orbit branching 
off (Miralda-Escud\'e \& Schwarzschild 1989).
In models without constant-density cores, the banana bifurcation 
can occur at quite small radii, and in models with strong cusps 
or central singularities, the long-axis orbit is unstable 
at all energies.
As Figure 2 shows, the long-axis orbit in triaxial models with 
Dehnen's density law is stable at most energies only when the 
cusp is weak, $\gamma \leqsim 0.7$, and the figure round, $c/a\geqsim 0.7$.
Both conditions are violated by the majority of low-luminosity 
ellipticals; thus, box orbits -- which require for their 
existence a stable long-axis orbit -- should not be present 
in most of these galaxies (even assuming that they do not contain nuclear
black holes).
Brighter ellipticals, which tend to be rounder and to have 
shallower cusps, may support box orbits, but 
only at small to intermediate radii where the long-axis orbit 
is stable, and only if they do not contain nuclear black holes.

\figureps[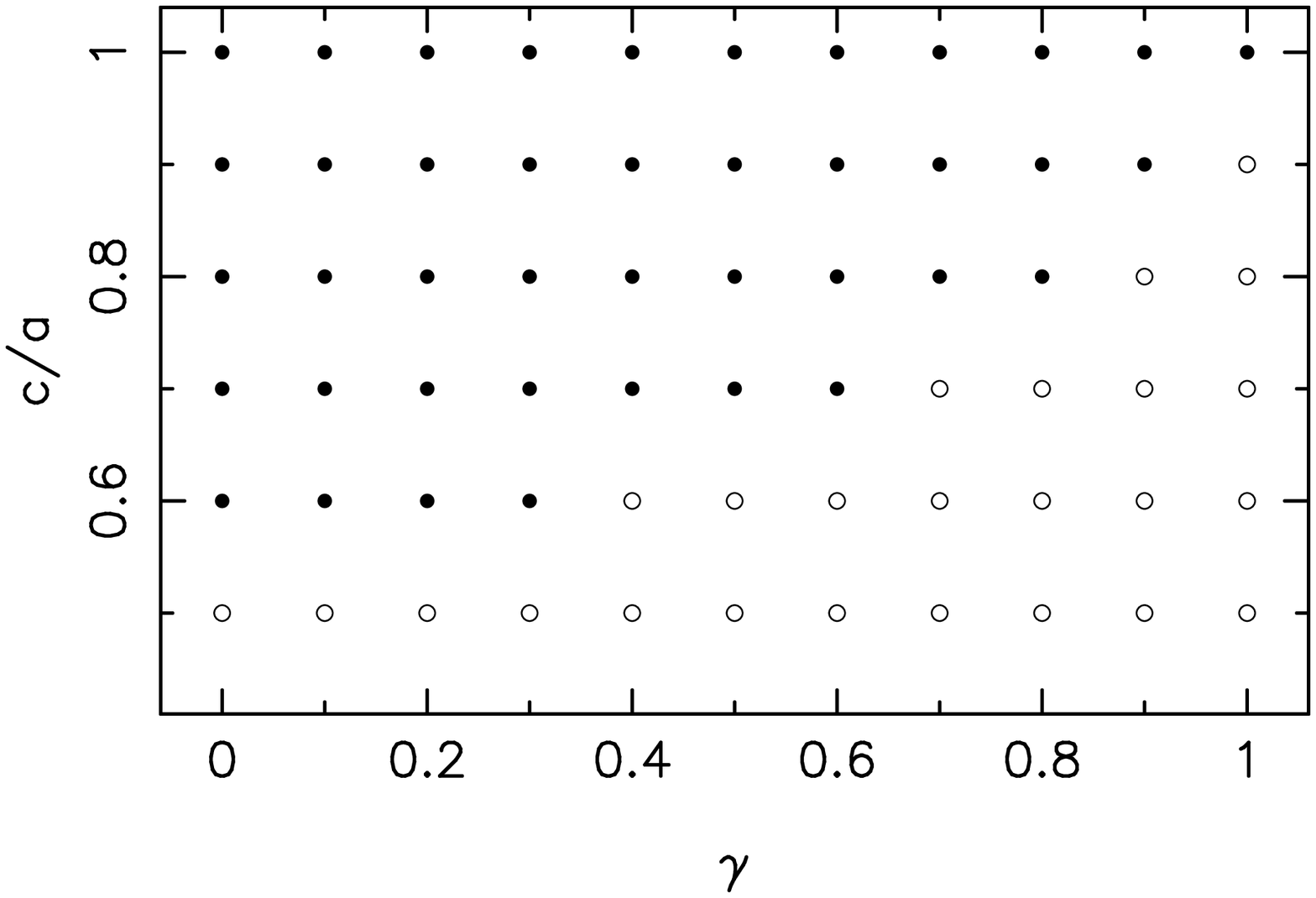,.6\hsize] 2. 
Stability of the long-axis orbit
in triaxial models with various short-to-long axis ratios $c/a$
and cusp slopes $\gamma$ (Fridman 1997).
A solid dot indicates stability of the long-axis orbit that just touches
the equipotential surface corresponding to the ellipsoidal shell that
divides the model into two regions of equal mass. 
Every model has ``maximal triaxiality,'' i.e. $(a^2-b^2)/(a^2-c^2)=0.5$.
For $\gamma>1$, the long-axis orbit is always unstable.

Box orbits are strongly populated in the self-consistent triaxial 
models with large cores (Schwarzschild 1979; Statler 1987).
The absence of bona-fide box orbits is not necessarily lethal
to the triaxial hypothesis, however, since higher-order resonances can also 
serve as the generators of boxlike orbits.
An obvious candidate is the 2:1 banana orbit that bifurcates 
from the unstable long-axis orbit.
However the range of shapes of the regular orbits associated 
with the 2:1 resonance is relatively small.
The reason is that the banana orbits pass quite near to the origin,
and even a modest perturbation is sufficient to drive them into 
the destabilizing center.
This means that the orbits from the 2:1 family
are not able to reproduce the shapes of the most 
highly elongated box orbits.
The problem becomes more severe as the elongation of the 
potential increases, since the bending angle of the banana orbits 
goes up as the potential becomes flatter (Pfenniger \& de Zeeuw 
1989).

Schwarzschild (1993) explored the degree to which regular orbits 
alone could reproduce the mass distribution of triaxial models 
with scale-free, $\rho\propto r^{-2}$ density laws.
He found that self-consistency could not be achieved for highly 
elongated triaxial models with $c/a = 0.3$.
Merritt \& Fridman (1996) found that they could not attain 
self-consistency using just the regular orbits in a model 
with Dehnen's density law, $\gamma=2$, $c/a=0.5$ and 
$T=(a^2-b^2)/(a^2-c^2)=0.5$.
Work in progress (Merritt 1997) suggests that the restriction to 
regular orbits in triaxial models with strong cusps limits the 
degree of triaxiality to values of $T$ less than
$\sim 0.3$ or greater than $\sim 0.8$.
Real galaxies must either avoid these shapes, or else they must 
incorporate stochastic orbits in order to achieve self-consistency.

\section Chaos and Mixing

Even Schwarzschild's first triaxial model contained a signficant 
number of sto-chastic orbits.
This was discovered (Merritt 1980) when the orbits were re-integrated 
using a different computer -- about 10\% of 
them generated different occupation numbers 
than in the original integrations, a result of 
the well-known sensitivity of stochastic orbits to small perturbations.
Orbital stochasticity becomes more important as the central concentration 
of a triaxial model is increased, since a small core radius or a 
strong cusp induces instability in orbits that pass near the 
center.
In the scale-free, $\rho\propto r^{-2}$ triaxial potentials investigated by 
Schwarzschild (1993), many or most of the orbits with boxlike 
(i.e. stationary) initial conditions were stochastic.
Merritt \& Valluri (1996) found the same to be true in triaxial 
potentials with the ``imperfect'' density law
$$
\rho(m) = {\rho_0\over (r_0^2+m^2)(1+m^2)},
\ \ \ \ m^2={x^2\over a^2} + {y^2\over b^2} + {z^2\over c^2}.
$$
For $r_0=1$, this reduces to the Perfect Ellipsoid (de Zeeuw \&
Lynden-Bell 1985), while for $r_0=0$ the model has a $\rho\propto 
m^{-2}$ 
central density cusp.
Even for $r_0=0.1$, the majority of boxlike orbits were found to 
be stochastic, and the fraction increased as $r_0$ was reduced to 
$0.01$ or $0.001$.
Adding a central point mass containing $\sim 0.3\%$ or more of 
the total mass was also found to be effective at destroying the regularity of 
most of the box orbits, consistent with the prediction of Gerhard 
\& Binney (1985).

Stochastic motion is qualitatively different from regular 
motion: a stochastic trajectory is not quasi-periodic, and perturbations 
of a stochastic orbit grow exponentially with time.
Until recently, model builders tended to ignore stochastic 
orbits, both because they were thought to be rare in triaxial 
potentials and because their shapes seemed to make them
poor building blocks for galaxies.
But lately the focus has changed.
While model builders may prefer to avoid stochastic orbits, 
there is no reason for nature to do so; the fraction of stars on 
stochastic orbits in a real galaxy is probably comparable to the 
fraction of phase space that is stochastic.
Furthermore, there is an interesting new timescale associated 
with stochasticity, the chaotic mixing timescale (Kandrup \& 
Mahon 1994).
An initially localized clump of stars in stochastic phase space 
will spread as the stellar trajectories diverge; the timescale 
for the divergence is initially equal to the inverse of the 
Liapunov exponent that characterizes the orbital instability.
Because stochastic motion is essentially random over long periods 
of time, the probability of finding a single star from the 
ensemble anywhere in stochastic phase space tends toward a constant; 
in other words, the distribution of stars evolves toward a steady state.
Something similar to this takes place in regular phase space, as the 
phases of stars on nearby orbits gradually move out of synch.
But chaotic mixing is often more efficient than phase mixing, 
since the region accessible to a stochastic orbit is much larger 
than the single torus to which a regular orbit is confined, and 
since the divergence in stochastic phase space grows exponentially
with time (Merritt 1996).

If chaotic mixing were always an efficient process, one would 
expect the stochastic parts of phase space to be fully mixed in 
real triaxial galaxies.
The net effect would be to remove many or most of the boxlike 
orbits from solution space, and to replace them with the much smaller 
set of ``orbits'' (i.e. invariant densities) associated with stochastic 
phase space at each energy.
However the timescale associated with chaotic mixing in triaxial 
galaxies is a strong function of the degree of central 
concentration.
Goodman \& Schwarzschild (1981) found that the stochastic orbits 
in a triaxial model with a large core behaved essentially like 
regular orbits for $\sim 10^2$ oscillations.
Merritt \& Valluri (1996) likewise found a slow rate of chaotic 
mixing in ``imperfect ellipsoids'' (Eq. 3) with $r_0=0.1$.
However when $r_0$ was reduced to $0.01$ or $0.001$, the 
mixing time dropped to a small multiple of the crossing time.
Many dynamical systems exhibit such a steep dependence of the
chaotic mixing rate on the parameters defining the potential 
(e.g. Contopoulos et al. 1996).

The strong dependence of the chaotic mixing rate on the degree of 
central mass concentration in triaxial models leads to an interesting 
prediction (Merritt \& Valluri 1996).
Low-luminosity ellipticals, which have strong cusps as well as 
short dynamical times, ought to be well-mixed; while bright 
ellipticals, which have shallower cusps and longer crossing 
times, need not be.
Faint ellipticals are therefore less likely than bright ones
to be strongly triaxial.
In fact there is a striking change in the apparent shape 
distribution of elliptical galaxies near $M_B=-20$ (Fig. 1); this 
is roughly the same magnitude at which the typical cusp slope 
changes from $\gamma\approx 2$ to $\gamma\approx 1$ (Gebhardt et 
al. 1996).
Tremblay \& Merritt (1996) found that the Hubble-type distribution 
for the faint (strong cusp) ellipticals could be well reproduced under the 
axisymmetric hypothesis, while that of the brighter ellipticals 
could not, suggesting that some of the the latter were triaxial.
On the other hand, if dynamically signficant black holes are 
components of most elliptical galaxies -- as they seem to be in 
M32 and M87 -- then chaotic mixing of the boxlike trajectories 
should be efficient regardless of cusp slope.
Whichever is the case, mixing will have proceeded 
farther in the central regions of elliptical galaxies than in the 
envelopes; thus axisymmetry should be most common at small radii.

\section Do Triaxial Galaxies Exist?

The triaxial hypothesis was strengthened early on 
by a series of observational studies that detected departures 
from axisymmetry in many early-type galaxies.
Isophote twists were widely interpreted as signatures of triaxiality 
(Benacchio \& Galletta 1980; Leach 1981).
But twists at large radii could also be intrinsic, resulting 
from tidal interactions or accretion in an inclined plane.
Many galaxies clearly fall into this category, e.g. Centaurus A.
And some galaxies with the strongest twists are probably not 
ellipticals at all.
Fasano \& Bonoli (1989) note that significant twisting is only 
seen in galaxies that deviate from a de Vaucouleurs luminosity 
profile, and suggest that some fraction of twisted ellipticals 
are barred S0's.
Nieto et al. (1992) note further that many galaxies with large twists 
exhibit strong changes in the isophote shapes at the radius where 
the position-angle change is the greatest.
They argue that most of these galaxies are barred S0's, and show that
both NGC 596 and 1549 -- two of the prototypical 
``twisted ellipticals'' -- have isophotal morphologies 
very similar to those of known SB0's.

Kinematical tests for triaxiality have also not fared well.
The same trick of projection that causes the 
isophotes of a triaxial system to twist can also induce a twist 
in the stellar velocity field (Contopoulos 1956; Binney 1985).
Franx et al. (1991) identified an unbiased sample of 38 
ellipticals with measured rotation curves along the major 
and minor axes and tested whether the distribution of kinematic 
misalignment angles, $\Psi = \tan^{-1}(v_{\rm minor}/v_{\rm major})$, 
was consistent with various hypotheses about the intrinsic 
shapes.
The found $\Psi$ to be strongly peaked around 
$\Psi=0$ (rotation about the apparent minor axis) and $\Psi=\pi/2$
(rotation around the apparent major axis), with most galaxies 
having $\Psi\approx 0$.
A much smaller fraction of galaxies have strong misalignments, 
$v_{\rm minor}\approx v_{\rm major}$.
While this distribution is not inconsistent with triaxiality, 
it is more naturally reproduced by assuming that $\sim 60\%$ of 
ellipticals are oblate and $\sim 40\%$ prolate, with perhaps a 
handful -- those with the strongest misalignments -- triaxial.

The handful of elliptical galaxies with strong kinematic misalignments 
are among the best current candidates for triaxiality.
Since they are so few in number, it makes sense to examine 
several of them in detail here.

{\bf NGC 1549} Franx et al. (1989) quote $v_{minor}/v_{major}=0.87$ 
for this E2 galaxy.
Malin \& Carter (1983) note the existence of faint shells 
and point out that NGC 1549 appears to be interacting with 
its neighbor NGC 1553.
Longo et al. (1994) present rotation curves along two position 
angles and stress the ``very strange kinematical behavior'':
the minor-axis rotation curve is U-shaped, a likely indicator 
of a recent interaction.
Nieto et al. (1992) argue that NGC 1549 is a misclasssified SB0.

{\bf NGC 2749} An E2 galaxy; Jedrzejewski \& Schechter (1989) find 
$v_{minor}/$ $v_{major}\approx1.1$.
But they note that its high major-axis rotation places this 
galaxy above the ``oblate isotropic rotator'' line, and argue 
that it may be a mis-classified S0.
They also find strong $5007$\AA\ emission in the inner regions 
suggesting the recent accretion of a gas-rich galaxy.

{\bf NGC 4365 and 4406} These E3 galaxies 
exhibit an extreme sort of kinematic misalignment:
the angular momenta of the central and outer regions are nearly orthogonal.
Both galaxies are minor-axis rotators at large radii, hence probably 
prolate, but the rotation near the center is along the major axis.
Such strong misalignments suggest that the core 
material was accreted (Kormendy 1984; Balcells \& Quinn 1990),
and that different orbital families are populated at large and small 
radii (Statler 1991).

{\bf NGC 4589} M\"ollenhoff \& Bender (1989) find 
$v_{minor}/v_{major}\approx 0.65$ for this E2 galaxy.
Statler (1994a) carried out a detailed analysis of NGC 4589 using 
the M\"ollenhoff \& Bender velocities and concluded
that the galaxy was significantly prolate-triaxial, with 
$T\approx 0.65$ and $c/a\approx 0.8$.
But M\"ollenhoff \& Bender note the ``fairly complex''
stellar rotation field of NGC 4589, including bumps in the 
major-axis rotation curve.
They argue for the recent accretion of a gas-rich 
companion in order to explain the prominent minor-axis
dust lane.

{\bf NGC 5128} Centaurus A is a nearby giant elliptical with a 
prominent dust lane and an extensive gas disk.
Evidence for strong departures from axisymmetry has been 
adduced from the gas motions (Graham 1979), the stellar velocity 
field (Wilkinson et al. 1986) and the kinematics of the planetary nebula 
system (Hui et al. 1995).
But this galaxy is almost certainly the product of a recent merger 
event (e.g. Schweizer 1987); the gas distribution in particular 
appears to be in a transient state (Bertola et al. 1985).

{\bf NGC 7145} An E0 galaxy; Franx et al. (1989) find 
$v_{minor}/v_{major}=0.72$.
It is a shell galaxy (Malin \& Carter 1983) and a member of a 
close pair.
Nieto et al. (1992) argue that this galaxy too is a misclassified 
SB0.

{\bf NGC 1700} This E3 galaxy with a strong cusp 
has been the subject of probably the most detailed kinematical 
study to date of a single elliptical.
Statler et al. (1996) mapped the stellar velocities out to almost five 
effective radii along four position angles.
The presence of shells (Forbes \& Thomson 1992) and strongly 
box-shaped, almost square isophotes at large radii (Franx et al. 1989) 
suggest that this galaxy experienced a significant merger or 
accretion event within the last several Gyr (Schweizer \& Seitzer 1992).
The rotational velocity field within $\sim 2.5R_e$ is symmetric, 
with the exception of a weakly counter-rotating (but aligned) 
core; the inner isophotes exhibit no twists.
Statler et al. found that an oblate model could reproduce 
these inner data extremely well.
Starting at $\sim 3R_e$, both the photometric and kinematic axes 
begin to twist; but beyond $\sim 4R_e$, the galaxy is 
clearly not in a dynamically relaxed state, since the velocities 
reverse along one axis producing a U-shaped rotation curve.
Statler et al. argued for a model in which the inner regions of 
NGC 1700 are relaxed and oblate, while the outer regions are 
still evolving in response to the accretion of a smaller galaxy.
The twists first occur at the radius of transition between the 
relaxed and evolving regions.

As these examples show, departures from axisymmetry are often 
accompanied by signatures of recent dynamical interactions and/or by 
hints that the observed galaxy is a barred S0.
While many early-type galaxies are clearly not axisymmetric,
it is less clear whether their figures would naturally be described 
as ``triaxial'' in the sense of stationary, nested ellipsoids.
By contrast, the case for axisymmetry is extremely good in a 
number of elliptical galaxies, including M32 (Qian et al. 1995) 
and NGC 3379 (Statler 1994b).
And in NGC 1700, which is strongly non-axisymmetric at large 
radii, the inner regions have apparently chosen to relax to an oblate shape.
These axisymmetric galaxies all have strong cusps, and M32 probably contains 
a massive nuclear black hole as well (Bender et al. 1996).
Thus there appears to be some support for the hypothesis that 
global shapes are correlated with the degree of central mass 
concentration.
But the overall evidence for {\it persistent} triaxiality in elliptical 
galaxies remains weak.

There is however another class of stellar system for which 
triaxiality seems to be increasingly implicated.
Kormendy (1982) has emphasized that the bulges of barred spiral 
galaxies often appear to be misaligned both with the bar and with the 
external disk, implying that the bulges are not axisymmetric.
This ``triaxial bulge'' or ``bar-within-a-bar'' phenomenon is 
now known to be quite common (Wozniak et al. 1995).
(With a few notable exceptions -- e.g. M31 (Stark 1977) -- the bulges of 
non-barred spirals appear to be accurately spheroidal.)
Again, it is not known whether these configurations are transient 
or long-lived, though $N$-body simulations suggest that 
multiply-barred systems can persist for many rotations 
(Sellwood \& Merritt 1994).
Nevertheless it is striking that triaxiality -- which was first put 
forward as a way of explaining the {\it slow} rotation of elliptical 
galaxies -- seems to find its most common expression in 
rapidly-rotating systems like bulges and bars.

\acknowl I thank T. Statler and M. Valluri for critical comments
on the manuscript. 

\references

Balcells, M. \& Quinn, P. J. 1990, ApJ, 361, 381

Benacchio, L. \& Galletta, G. 1980, MNRAS, 193, 885

Bender, R., Kormendy, J. \& Dehnen, W. 1996, preprint

Bertola, F., Galletta, G. \& Zeilinger, W. W. 1985, ApJ 292, 
L51

Binney, J. 1978, Comm Ap, 8, 27

Binney, J. 1985, MNRAS, 212, 767

Contopoulos, G. 1956, Z Ap 39, 126

Contopoulos, G., Voglis, N. \& Efthymiopoulos, C. 1996, in Nobel 
Symposium 98, Barred Galaxies and Circumnuclear Activity

Dehnen, W. 1993, MNRAS, 265, 250

de Zeeuw, P. T. 1985, MNRAS, 216, 273

de Zeeuw, P. T. and Lynden-Bell, D. 1985, MNRAS, 215, 713

Fasano, G. \& Bonoli, C. 1989, AAp S 79, 291

Ferrarese, L., van den Bosch, F. C., Ford, H. C., 
Jaffe, W. \& O'Connell, R. W. 1994, AJ, 108, 1598

Forbes, D. A. \& Thomson, R. C. 1992, MNRAS, 254, 723

Franx, M., Illingworth, G. \& Heckman, T. 1989, ApJ, 344, 613

Franx, M., Illingworth, G. D. \& de Zeeuw, P. T. 1991, ApJ, 383, 
112

Fridman, T. 1997, PhD Thesis, Rutgers University

Gebhardt, K. et al. 1996, AJ, 112, 105

Gerhard, O. \& Binney, J. 1985, MNRAS, 216, 467

Goodman, J. \& Schwarzschild, M. 1981, ApJ, 245, 1087

Graham, J. A. 1979, ApJ, 232, 60

Heiligman, G. \& Schwarzschild, M. 1979, ApJ, 233, 872

Hui, X., Ford, H. C., Freeman, D. C. \& Dopita, M. A. 1995, ApJ, 449, 592

Jedrzejewski, R. \& Schechter, P. L. 1989, AJ, 98, 147

Kandrup, H. E. and Mahon, M. E. 1994, Phys Rev E, 49, 3735

Kormendy, J. 1982, ApJ, 257, 75

Kormendy, J. 1984, ApJ, 287, 577

Kormendy, J. \& Richstone, D. O. 1995, AAR\&A, 33, 581

Kuzmin, G. G. 1973, in Dynamics of Galaxies and Clusters, ed. T. B. Omarov 
(Alma Ata: Akad. Nauk. Kaz. SSR), 71

Lauer, T. R. et al. 1995, AJ, 110, 2622 

Leach, R. 1981, ApJ, 248, 485

Longo, G., Zaggia, S. R., Busarello, G. \& Richter, G. 1994, AAp S, 
105, 433

Malin, D. F. \& Carter, D. 1983, ApJ, 274, 534

Merrifield, M. \& Kuijken, K. 1992, ApJ, 432, 575

Merritt, D. 1980, ApJ S, 43, 435

Merritt, D, 1996, Cel. Mech., 64, 55

Merritt, D. 1997, in preparation

Merritt, D. \& de Zeeuw, T. 1982, ApJ, 267, L19

Merritt, D. \& Fridman, T. 1995, ASP Conf. Ser. Vol. 86, 
Fresh Views of Elliptical Galaxies, ed. A. Buzzoni et al. 
(Provo: ASP), 13

Merritt, D. and Fridman, T. 1996, ApJ, 460, 136

Merritt, D. \& Valluri, M. 1996, ApJ, 471, 82

Miralda-Escud\'e, J. \& Schwarzschild, M. 1989, ApJ, 339, 752

M\"ollenhoff, C. \& Bender, R. 1989, AAp, 214, 61

Nieto, J.-L., Bender, R., Poulain, P. \& Surma, P. 1992, AAp, 
257, 97

Norman, C., May, A. \& van Albada, T. 1985, ApJ, 296, 20

Pfenniger, D. \& de Zeeuw, T. 1989, in Dynamics of Dense Stellar 
Systems, ed. D. Merritt (Cambridge: CUP), 81

Qian, E. E., de Zeeuw, P. T., van der Marel, R. P. \& Hunter, C. 
1995, MNRAS, 602, 622

Richstone, D. O. \& Tremaine, S. 1985, ApJ, 296, 370

Rubin, V., Graham, J. \& Kenney, J. 1992, ApJ, 394, L9

Schwarzschild, M. 1979, ApJ, 232, 236

Schwarzschild, M. 1982, ApJ, 263, 599

Schwarzschild, M. 1993, ApJ, 409, 563

Schweizer, F. 1987, in IAU Symp. No. 127, Structure and Dynamics 
of Elliptical Galaxies, ed. T. de Zeeuw (Reidel: Dordrecht), 109

Schweizer, F. \& Seitzer, P. 1992, AJ, 104, 1039

Sellwood, J. A. \& Merritt, D. 1994, ApJ, 425, 530

Stark, A. A. 1977, ApJ, 213, 368

Statler, T. 1987, ApJ, 321, 113

Statler, T. 1991, AJ, 102, 882

Statler, T. 1994a, ApJ, 425, 500

Statler, T. 1994b, AJ, 108, 111

Statler, T., Smecker-Hane, T. \& Cecil, G. N. 1996, AJ, 111, 1512

Tremblay, B. \& Merritt, D. 1996, AJ, 111, 2243

Wilkinson, A., Sharples, R. M., Fosbury, R. A. E. \& Wallace, P. 
T. 1986, MNRAS, 218, 297

Wozniak, H., Friedli, D., Martinet, L., Martin, P. \& Bratschi, 
P. 1995, AAp S, 111, 115

Young, P. J., Westphal, J. A., Kristian, J., Wilson, C. P. 
\& Landauer, F. P. 1978, ApJ, 221, 721

\bye